\def\IEEEsubmission{0}

\def\complexNumbers{\mathbb{C}}

\def\constante{{\rm e}}
\def\constantj{{\rm j}}

\def\transmittedSignal[#1]{p\left(#1\right)}
\def\numberOfOccupiedSubcarriers{D}
\def\frequencyDeviation{\Delta f}
\def\numberOfShifts{M}
\def\indexSubcarrier{k}
\def\indexTime{m}
\def\basisFunction[#1]{B_{#1}(\timeVar)}
\def\amountOfShift[#1]{\tau_{#1}}
\def\dataSymbols[#1]{d_{#1}}
\def\angleSignal[#1]{\psi_{#1}(\timeVar)}
\def\instantaneousFrequency[#1]{F_{#1}(t)}
\def\besselFunctionFirstKind[#1][#2]{J_{#1}\left(#2\right)}
\def\lowerFrequency{L_{\rm d}}
\def\upperFrequency{L_{\rm u}}
\def\idftSize{N}
\def\indexSample{n}
\def\fourierSeries[#1]{c_{#1}}

\def\CPDuration{T_{\rm CP}}
\def\sampleDuration{T_{\rm sample}}
\def\symbolDuration{T_{\rm s}}
\def\timeVar{t}

\def\fresnelC[#1]{C(#1)}
\def\fresnelS[#1]{S(#1)}
\def\linearXone{x_1}
\def\linearXtwo{x_2}

\def\fcarrier{f_{\rm c}}

\newcommand\mydots{\hbox to 1em{.\hss.\hss.}}

\if \IEEEsubmission0
	\documentclass[journal]{IEEEtran}
\else
	\documentclass[journal,onecolumn,draftclsnofoot]{IEEEtran} 
\fi

\usepackage{multicol}
\usepackage{lipsum}
\usepackage{mathtools}
\usepackage{amsthm}
\usepackage{acronym}
\usepackage[bookmarksopen=true]{hyperref}
\usepackage{stfloats}
\usepackage{amsfonts}
\usepackage{acronym}
\usepackage{cite}
\usepackage{multirow}
\usepackage{bm}
\usepackage{amsmath,amssymb}
\usepackage{graphicx}
\usepackage[table,xcdraw]{xcolor}
\usepackage[english]{babel}
\usepackage[caption=false,font=footnotesize]{subfig}
\usepackage[geometry]{ifsym}
\usepackage{array}
\usepackage[utf8]{inputenc}
\usepackage[T1]{fontenc}

\usepackage{tikz}
\usepackage{lipsum}
\usetikzlibrary{calc,shadings,patterns}
\makeatletter
\tikzset{%
  remember picture with id/.style={%
    remember picture,
    overlay,
    save picture id=#1,
  },
  save picture id/.code={%
    \edef\pgf@temp{#1}%
    \immediate\write\pgfutil@auxout{%
      \noexpand\savepointas{\pgf@temp}{\pgfpictureid}}%
  },
  if picture id/.code args={#1#2#3}{%
    \@ifundefined{save@pt@#1}{%
      \pgfkeysalso{#3}%
    }{
      \pgfkeysalso{#2}%
    }
  }
}

\def\savepointas#1#2{%
  \expandafter\gdef\csname save@pt@#1\endcsname{#2}%
}

\def\tmk@labeldef#1,#2\@nil{%
  \def\tmk@label{#1}%
  \def\tmk@def{#2}%
}

\tikzdeclarecoordinatesystem{pic}{%
  \pgfutil@in@,{#1}%
  \ifpgfutil@in@%
    \tmk@labeldef#1\@nil
  \else
    \tmk@labeldef#1,(0pt,0pt)\@nil
  \fi
  \@ifundefined{save@pt@\tmk@label}{%
    \tikz@scan@one@point\pgfutil@firstofone\tmk@def
  }{%
  \pgfsys@getposition{\csname save@pt@\tmk@label\endcsname}\save@orig@pic%
  \pgfsys@getposition{\pgfpictureid}\save@this@pic%
  \pgf@process{\pgfpointorigin\save@this@pic}%
  \pgf@xa=\pgf@x
  \pgf@ya=\pgf@y
  \pgf@process{\pgfpointorigin\save@orig@pic}%
  \advance\pgf@x by -\pgf@xa
  \advance\pgf@y by -\pgf@ya
  }%
}

\makeatother

\newcounter{hatchNumber}
\DeclarePairedDelimiter\floor{\lfloor}{\rfloor}

\usepackage{cuted}
\makeatletter
\newif\ifAC@uppercase@first%
\def\Aclp#1{\AC@uppercase@firsttrue\aclp{#1}\AC@uppercase@firstfalse}%
\def\AC@aclp#1{%
	\ifcsname fn@#1@PL\endcsname%
	\ifAC@uppercase@first%
	\expandafter\expandafter\expandafter\MakeUppercase\csname fn@#1@PL\endcsname%
	\else%
	\csname fn@#1@PL\endcsname%
	\fi%
	\else%
	\AC@acl{#1}s%
	\fi%
}%
\def\Acp#1{\AC@uppercase@firsttrue\acp{#1}\AC@uppercase@firstfalse}%
\def\AC@acp#1{%
	\ifcsname fn@#1@PL\endcsname%
	\ifAC@uppercase@first%
	\expandafter\expandafter\expandafter\MakeUppercase\csname fn@#1@PL\endcsname%
	\else%
	\csname fn@#1@PL\endcsname%
	\fi%
	\else%
	\AC@ac{#1}s%
	\fi%
}%
\def\Acfp#1{\AC@uppercase@firsttrue\acfp{#1}\AC@uppercase@firstfalse}%
\def\AC@acfp#1{%
	\ifcsname fn@#1@PL\endcsname%
	\ifAC@uppercase@first%
	\expandafter\expandafter\expandafter\MakeUppercase\csname fn@#1@PL\endcsname%
	\else%
	\csname fn@#1@PL\endcsname%
	\fi%
	\else%
	\AC@acf{#1}s%
	\fi%
}%
\def\Acsp#1{\AC@uppercase@firsttrue\acsp{#1}\AC@uppercase@firstfalse}%
\def\AC@acsp#1{%
	\ifcsname fn@#1@PL\endcsname%
	\ifAC@uppercase@first%
	\expandafter\expandafter\expandafter\MakeUppercase\csname fn@#1@PL\endcsname%
	\else%
	\csname fn@#1@PL\endcsname%
	\fi%
	\else%
	\AC@acs{#1}s%
	\fi%
}%
\edef\AC@uppercase@write{\string\ifAC@uppercase@first\string\expandafter\string\MakeUppercase\string\fi\space}%
\def\AC@acrodef#1[#2]#3{%
	\@bsphack%
	\protected@write\@auxout{}{%
		\string\newacro{#1}[#2]{\AC@uppercase@write #3}%
	}\@esphack%
}%
\def\Acl#1{\AC@uppercase@firsttrue\acl{#1}\AC@uppercase@firstfalse}
\def\Acf#1{\AC@uppercase@firsttrue\acf{#1}\AC@uppercase@firstfalse}
\def\Ac#1{\AC@uppercase@firsttrue\ac{#1}\AC@uppercase@firstfalse}
\def\Acs#1{\AC@uppercase@firsttrue\acs{#1}\AC@uppercase@firstfalse}

\acrodef{SIC}{successive interference cancellation}
\acrodef{PAPR}{peak-to-average-power ratio}
\acrodef{APAC}{aperiodic autocorrelation}
\acrodef{OFDM}{orthogonal frequency division multiplexing}
\acrodef{DFT}{discrete Fourier transform}
\acrodef{DC}{direct current}
\acrodef{CS}{complementary sequence}
\acrodef{GCP}{Golay complementary pair}
\acrodef{ANF}{algebraic normal form}
\acrodef{PSK}{phase shift keying}
\acrodef{QAM}{quadrature amplitude modulation}
\acrodef{QPSK}{quadrature phase shift keying}
\acrodef{GDJ}{Golay-Davis-Jedwab}
\acrodef{PMEPR}{peak-to-mean envelope power ratios}
\acrodef{FFT}{fast Fourier transform}
\acrodef{BER}{bit-error ratio}
\acrodef{SNR}{signal-to-noise ratio}
\acrodef{4G}{Fourth Generation}
\acrodef{5G}{Fifth Generation}
\acrodef{NR}{New Radio}
\acrodef{LTE}{Long-Term Evolution}
\acrodef{PTS}{partial transmit sequences}
\acrodef{PSD}{power spectral density}
\acrodef{LDPC}{low-density parity check}
\acrodef{SE}{spectral efficiency}
\acrodef{eLAA}{enhanced licensed-assisted access}
\acrodef{NR-U}{NR-Unlicensed}
\acrodef{RM}{Reed-Muller}
\acrodef{AE}{autoencoder}
\acrodef{DNN}{deep neural network}
\acrodef{OFDM-AE}{OFDM-based autoencoder}
\acrodef{DL}{deep learning}
\acrodef{CP}{cyclic prefix}
\acrodef{AWGN}{additive white Gaussian noise}
\acrodef{P2C}{polar-to-Cartesian}
\acrodef{CFR}{channel frequency response}
\acrodef{ReLU}{rectified linear unit}
\acrodef{MMSE}{minimum mean sqaure error}
\acrodef{BPSK}{binary phase shift keying}
\acrodef{BLER}{block error rate}
\acrodef{ML}{machine learning}
\acrodef{PHY}{physical layer}
\acrodef{PA}{power amplifier}
\acrodef{IDFT}{inverse DFT}
\acrodef{DoF}{degrees-of-freedom}
\acrodef{IoT}{Internet-of-Things}
\acrodef{DFT-s-OFDM}{discrete Fourier transform-spread orthogonal frequency division multiplexing}
\acrodef{MMSE}{minimum mean square error}
\acrodef{FDE}{frequency-domain equalization}
\acrodef{FrFT}{fractional Fourier transform}
\acrodef{TF}{time-frequency}
\acrodef{BFSK}{binary frequency-shift keying}
\acrodef{CSS}{chirp spread spectrum}
\acrodef{BCSS}{binary chirp spread spectrum}
\acrodef{EVA}{Extended Vehicular A}
\acrodef{MIMO}{multi-input multi-output}
\acrodef{PIC}{parallel interference cancellation}
\acrodef{LoRa}{Long Range}
\acrodef{HF}{high-frequency}
\acrodef{FDSS}{frequency-domain spectral shaping}
\acrodef{CSC}{circularly-shifted chirp}
\acrodef{ISI}{inter-symbol interference}
\acrodef{DFRC}{dual-function radar and communication}
\begin{document}
\title{ 
DFT-spread-OFDM Based Chirp Transmission
}
\author{Alphan~\c{S}ahin, Nozhan~Hosseini, Hosseinali~Jamal, Safi Shams Muhtasimul Hoque, David~W.~Matolak
\thanks{The authors are with the University of South Carolina, Columbia, SC. E-mail: asahin@mailbox.sc.edu, nozhan@email.sc.edu, hjamal@email.sc.edu, shoque@email.sc.edu,  matolak@cec.sc.edu. 

\cite{Safi_2020_CCNC} and \cite{Safi_2020_GC}  utilize the proposed framework in this manuscript. They are cited in this study as per IEEE Communications Letter submission policy.}
}


\maketitle

\begin{abstract}
  
In this study, we propose a framework for chirp-based communications by exploiting \ac{DFT-s-OFDM}. We show that a well-designed \ac{FDSS} filter for \ac{DFT-s-OFDM} can convert its single-carrier nature to a linear combination of chirps circularly translated in the time domain. Also, by exploiting the properties of the Fourier series and Bessel function of the first kind, we analytically obtain the \ac{FDSS} filter for an arbitrary chirp. We theoretically show that the chirps with low ripples in the frequency domain result in a lower \ac{BER} via less noise enhancement. We also address the noise enhancement by exploiting the repetitions in the frequency. The proposed framework offers a new way to efficiently synthesize chirps that can be used in \ac{IoT}, \ac{DFRC} or wireless sensing applications with  existing \ac{DFT-s-OFDM} transceivers.
\end{abstract}

\acresetall

\section{Introduction}
Chirps are prominent for radar and communication applications in the sense that they can sweep a large bandwidth while still being constant-envelope signals, i.e., they provide robustness against non-linear distortions. 
Recently, there is a growing interest in their potential use in today's major communication systems, e.g., 3GPP \ac{5G} \ac{NR}, \ac{LTE}, and IEEE 802.11 Wi-Fi,  to address some emerging applications such as short-range wireless sensing, \ac{DFRC}, and \ac{IoT}. However, the physical layers of these communication systems are based on \ac{OFDM}. In this study, we aim at addressing the challenge of synthesizing chirps within an \ac{OFDM} framework.


There is an extensive literature dealing with chirps for communications. To provide a longer range and higher resolution for radar systems, chirps were first utilized by S. Darlington at Bell Labs \cite{darlingtonPatent1949}. They also introduce the idea of encoding bits as negative or positive slopes in \ac{TF} plane. In \cite{nozh2},  an orthogonal amplitude-variant linear chirp set where each chirp has a different chirp rate is investigated. In \cite{fresnel_nozh_7523229}, X. Ouyang and J. Zhao presented a way of constructing orthogonal chirps by introducing a term to the exponent of \ac{DFT} kernels. As the proposed method in \cite{fresnel_nozh_7523229} translates the chirps in the frequency domain, the signal bandwidth increases with the number of chirps. To limit the bandwidth, they proposed additional up-sampling and filtering operations, which folds the chirps in the frequency domain. The Fresnel transform and \ac{FrFT} were adopted to generate orthogonal chirp sets in \cite{fresnel_nozh_7523229} and \cite{fractional_for_nozh}, respectively. 
In \cite{Receiver_design_nozh}, an iterative receiver is proposed to decrease the \ac{BER} under frequency-selective channels for chirps. In addition, a proprietary \ac{CSS} modulation, called \ac{LoRa}, was introduced by Samtec for \ac{IoT} applications. The chirps for wireless sensing were also brought up in IEEE 802.11 Wi-Fi meetings, e.g., \cite{Xiao_huaweu_2019}. To the best of our knowledge, there is no study that investigates \acp{CSC} for chirp-based communications and reveals its relation to \ac{OFDM}. 

In this paper, we present a method that generates modulated \acp{CSC} based on \ac{DFT-s-OFDM} adopted in 3GPP \ac{5G}  \ac{NR} and 3GPP \ac{LTE}. The contributions of this study based on this method can be listed as follows:

{\bf A flexible framework for chirps}: The proposed framework relies on the design of \ac{FDSS} filter applied after  \ac{DFT} spreading to convert the Dirichlet sinc functions \cite{Sahin_2016} to a set of chirps translated uniformly in time. Since the introduced method controls the frequency domain behavior of the chirps over the \ac{OFDM} subcarriers, it allows utilization of techniques developed for \ac{OFDM}  for chirp-based communications. 

{\bf Theoretical FDSS design:} By exploiting the properties of the Bessel functions of the first kind, we show that the \ac{FDSS} filter for arbitrary chirps can be designed analytically based on the convolutions of up-sampled sequences relying on the Bessel function of the first kind. We also discuss the closed-form expressions for linear and sinusoidal chirps.

{\bf Transceiver compatibility:}
The proposed framework shows that existing \ac{DFT-s-OFDM} transceivers  can modulate and demodulate the \acp{CSC}, which enlarges the applications of existing \ac{DFT-s-OFDM} transceivers. We also  investigate the noise enhancement problem and its mitigation. 

{\bf Precise bandwidth control:} Since \acp{CSC} are generated in the frequency domain through an \ac{FDSS} filter, the proposed method achieves precise control of the chirp bandwidth without additional filtering operations mentioned in \cite{fresnel_nozh_7523229}.

\section{Circularly-Shifted Chirps with DFT-s-OFDM}
\begin{figure}[t]
	\centering
	{\includegraphics[width =3.3in]{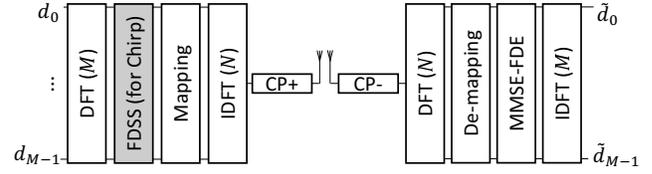}
	}
	\caption{
A DFT-s-OFDM transmitter  can synthesize modulated \acp{CSC} with a special FDSS filter designed based on the trajectory of chirp in time and frequency plane. A typical DFT-s-OFDM receiver can demodulate the modulated \acp{CSC} as the FDE removes the impact of FDSS on the subcarriers.
}
	\label{fig:blockdiagram}
\end{figure}

Consider a communication scheme where the data symbols are transmitted over the basis functions $\basisFunction[{\amountOfShift[0]}], \basisFunction[{\amountOfShift[1]}],$ $\mydots, \basisFunction[{\amountOfShift[\numberOfShifts-1]}]$ constructed by translating a chirp signal circularly in time, where $\amountOfShift[\indexTime]$ is the amount of circular shift. We assume that the shifts in time are uniformly spaced between 0 and $\symbolDuration$, i.e., $\amountOfShift[\indexTime]=\indexTime/\numberOfShifts\times\symbolDuration$, where $\symbolDuration$ is the chirp duration.
The complex baseband signal $\transmittedSignal[\timeVar]$ can then be expressed as 
\begin{align}
\transmittedSignal[\timeVar] = \sum_{\indexTime=0}^{\numberOfShifts-1} \dataSymbols[\indexTime] \basisFunction[{\amountOfShift[\indexTime]}]
= \sum_{\indexTime=0}^{\numberOfShifts-1} \dataSymbols[\indexTime] \constante^{\constantj\angleSignal[\indexTime]}
~,
\label{eq:originalWaveform}
\end{align}
where  $\dataSymbols[\indexTime]\in\complexNumbers$ is the $\indexTime$th data symbol (e.g., \ac{QPSK} symbols) and
 $\angleSignal[\indexTime]$ is the phase of the carrier for the $\indexTime$th basis function.  Therefore, the instantaneous frequency of the $\indexTime$th chirp signal $\basisFunction[{\amountOfShift[\indexTime]}]$ around the carrier frequency $\fcarrier$ can be obtained as $\instantaneousFrequency[\indexTime]=\frac{1}{2\pi}d{\angleSignal[\indexTime]}/{d\timeVar}$ Hz.
The waveform given in \eqref{eq:originalWaveform} is a  linear combination of the basis functions related to chirps. However, it is implicitly related to  \ac{DFT-s-OFDM}.  This relation can be shown as follows:

Let $\basisFunction[{\amountOfShift[\indexTime]}]=\constante^{\constantj\angleSignal[\indexTime]}$ be an  arbitrary band-limited function with the period of $\symbolDuration$. Hence, it can be expressed as
\begin{align}
\constante^{\constantj\angleSignal[\indexTime]} = \sum_{\indexSubcarrier=-\infty}^{\infty} \fourierSeries[\indexSubcarrier] \constante^{\constantj2\pi\indexSubcarrier\frac{\timeVar-\amountOfShift[\indexTime]}{\symbolDuration}}~,
\label{eq:basisDecompose}
\end{align}
where $\fourierSeries[\indexSubcarrier]$ is the $\indexSubcarrier$th Fourier coefficient given by
\begin{align}
\fourierSeries[\indexSubcarrier] = \frac{1}{\symbolDuration}\int_{\symbolDuration}
\constante^{\constantj\angleSignal[0]}
\constante^{-\constantj2\pi\indexSubcarrier\frac{\timeVar}{\symbolDuration}}d\timeVar. 
\end{align}
By using \eqref{eq:basisDecompose} and $\amountOfShift[\indexTime]=\indexTime/\numberOfShifts\times\symbolDuration$, \eqref{eq:originalWaveform} can be expressed as
\begin{align}
\transmittedSignal[\timeVar] 
\approx&\sum_{\indexSubcarrier=\lowerFrequency}^{\upperFrequency}\fourierSeries[\indexSubcarrier] \sum_{\indexTime=0}^{\numberOfShifts-1} \dataSymbols[\indexTime] 
\constante^{-\constantj2\pi \indexSubcarrier \frac{\indexTime}{\numberOfShifts}}
\constante^{\constantj2\pi \indexSubcarrier \frac{\timeVar}{\symbolDuration}}~,
\end{align}
where $\lowerFrequency<0$ and $\upperFrequency>0$ are integer values. The approximation is due to the fact that $\basisFunction[{\amountOfShift[\indexTime]}]$ is a band-limited function, i.e.,   $\fourierSeries[\indexSubcarrier]$ is a decaying Hermitian symmetric function as $\indexSubcarrier$ goes to positive or negative infinity. Finally, by sampling $\transmittedSignal[\timeVar]$ with the period  $\sampleDuration=\symbolDuration/\idftSize$, the discrete-time signal can be obtained as
\begin{align}
\transmittedSignal[\frac{\indexSample\symbolDuration}{\idftSize}] \approx&\underbrace{\sum_{\indexSubcarrier=\lowerFrequency}^{\upperFrequency}\underbrace{\fourierSeries[\indexSubcarrier]\underbrace{\sum_{\indexTime=0}^{\numberOfShifts-1} \dataSymbols[\indexTime] 
			\constante^{-\constantj2\pi \indexSubcarrier \frac{\indexTime}{\numberOfShifts}}}_{\numberOfShifts\text{-point DFT}}}_{\text{Frequency-domain spectral shaping}}
	\constante^{\constantj2\pi \indexSubcarrier \frac{\indexSample}{\idftSize}}}_{\idftSize\text{-point IDFT}}~.
\label{eq:chirpWave}
\end{align}
Hence, as shown in \figurename~\ref{fig:blockdiagram},  \eqref{eq:originalWaveform} is a special \ac{DFT-s-OFDM} symbol that can be implemented by 1) calculating the $\numberOfShifts$-point \ac{DFT} of a data vector, i.e., $[\dataSymbols[0], \dataSymbols[1],\mydots, \dataSymbols[\numberOfShifts-1]]$, 2) multiplying each element of the output of \ac{DFT} with the corresponding Fourier coefficient, i.e., \ac{FDSS} or windowing in frequency, and 3) calculating the $\idftSize$-point \ac{IDFT} of the shaped sequence after padding it with $\idftSize-(\upperFrequency-\lowerFrequency+1)$ zero symbols (e.g., guard subcarriers in \ac{OFDM}). 

In this study, without loss of generality, we assume $\lowerFrequency=\floor{\numberOfShifts/2}-\numberOfShifts+1$ and $\upperFrequency=\floor{\numberOfShifts/2}$ to ensure that the \ac{FDSS} occurs within the bandwidth spanned by $\numberOfShifts$ subcarriers. Therefore, the  chirp bandwidth should be less than or equal to $\numberOfShifts/\symbolDuration$. 
Let $\frequencyDeviation\triangleq\numberOfOccupiedSubcarriers/2\symbolDuration$ Hz be the maximum frequency deviation of each basis function, where $\numberOfOccupiedSubcarriers$ is a positive real number. Hence, the effective bandwidth of the transmitted signal $\transmittedSignal[\timeVar]$ is  $\numberOfOccupiedSubcarriers/\symbolDuration$ Hz. Therefore, $\numberOfOccupiedSubcarriers\le\numberOfShifts$ must hold true to form the \acp{CSC} via \eqref{eq:chirpWave}.

Note that \ac{FDSS} was discussed in 3GPP \ac{LTE} and 3GPP \ac{5G} \ac{NR} uplink as an implementation-specific option to reduce \ac{PAPR} further for \ac{DFT-s-OFDM}. Based on \eqref{eq:chirpWave}, we show the same \ac{DFT-s-OFDM} transmitter can also generate \acp{CSC} by choosing the shaping coefficients properly without compromising the other features of the physical layer design in these communication systems. As compared to the method in \cite{fresnel_nozh_7523229}, it also does not cause any bandwidth expansion as the chirps are circularly-shifted versions of each other in the time domain, which eliminates additional processing to avoid aliases discussed in \cite{fresnel_nozh_7523229}.

\def\SNR{{{SNR}}}
\def\SNRpost{{{SNR}_{\rm post}}}
\def\EBNo{E_{\rm b}/N_0}
\def\numberOfUtilizedBins{M_{\rm active}}
\def\spectralEfficiency{\rho}
\def\spectralEfficiencyPost{\rho_{\rm post}}
\def\signalFactor{{{\alpha}_{\rm MMSE}}}
\def\upsampleFactor{R}
\def\copyIndex{u}
\def\receivedSubcarrier[#1]{r_{#1}}
\def\receivedSubcarrierCombined[#1]{r'_{#1}}
\def\FDSScoefnew[#1]{c'_{#1}}
\subsection{Receiver}
\label{subsec:receiver}
As shown in \figurename~\ref{fig:blockdiagram}, a typical \ac{DFT-s-OFDM} receiver can demodulate the received \acp{CSC}. After the \ac{CP} is discarded, the \ac{DFT} of the received signal is calculated. In the frequency domain, the impact of the channel is removed with a single-tap \ac{MMSE} \ac{FDE}\footnote{Single-tap MMSE-FDE and linear MMSE estimator based on matrix notations of the corresponding operations are identical.}. The modulation symbols are obtained after an $\numberOfShifts$-\ac{IDFT} operation on the equalized signal vector. For a practical receiver, the shaping coefficients can be considered as part of the channel frequency response and estimated through channel estimation procedure. From this aspect, the proposed scheme does not require any change from a practical DFT-s-OFDM receiver. On the other hand, if the shaping coefficients are available at the receiver {\em a priori}, the receiver can perform better as they do not need to be estimated. 

It is worth noting that \ac{FDSS}  coefficients for chirps are often not unimodular, i.e., $|\fourierSeries[\indexSubcarrier]|\neq|\fourierSeries[l]|$ for all $\indexSubcarrier$ and $l$. Therefore, a single-tap \ac{FDE} causes noise enhancement for the subcarriers where $|\fourierSeries[\indexSubcarrier]|<1$. Also, \ac{ISI} occurs after the \ac{IDFT} de-spreading operation since \ac{MMSE} is a biased estimator. By utilizing the error rate analysis given for plain \ac{DFT-s-OFDM} \cite{Nisar_2007}, the overall impact of \ac{MMSE}-\ac{FDE} on  the data symbols can be calculated by obtaining the \ac{SNR}  after the equalization as
\begin{align}
\SNRpost = \frac{1}{\sqrt{1/\signalFactor} -1}~,
\label{eq:SNRpost}
\end{align}
where $\signalFactor=\left(1/\numberOfShifts\sum_{\indexSubcarrier=\lowerFrequency}^{\upperFrequency} \frac{|\fourierSeries[\indexSubcarrier]|^2}{|\fourierSeries[\indexSubcarrier]|^2+1/\SNR}\right)^2$. The equation \eqref{eq:SNRpost} shows that $\SNRpost\le\SNR$ and the equality holds for $|\fourierSeries[\indexSubcarrier]|=1$ for all $\indexSubcarrier$.   Under the assumption of the \ac{ISI} being Gaussian, the error rate for \acp{CSC} with arbitrary modulation can be calculated by using $\SNRpost$. As demonstrated in Section~\ref{sec:results} through theoretical results based on \eqref{eq:SNRpost}, this assumption is fairly accurate.

\subsubsection{Reducing Noise Enhancement} 
\label{subsubsec:noiseEnc}
The noise enhancement due to the \ac{FDSS} can be substantially reduced by achieving a frequency diversity {\em without} distorting transmitting \acp{CSC} as follows:
Assume that only every $\upsampleFactor-1$ other bins of the input of $\numberOfShifts$-point \ac{DFT} are utilized for data symbols. It can be shown that the resulting vector after the $\numberOfShifts$-point \ac{DFT} consists of $\upsampleFactor$ repetitions of the $\numberOfShifts/\upsampleFactor$-point \ac{DFT} of the data symbols. Since each repetition is multiplied with different \ac{FDSS} coefficients, it enables utilization of frequency diversity techniques without changing the original signal bandwidth. At the receiver side, $\upsampleFactor$ repetitions can be combined as $\receivedSubcarrierCombined[\indexSubcarrier]=\sum_{\copyIndex=0}^{\upsampleFactor-1} \fourierSeries[\indexSubcarrier+\copyIndex\numberOfShifts/\upsampleFactor]^*\receivedSubcarrier[{\indexSubcarrier+\copyIndex\numberOfShifts/\upsampleFactor}]$ for $\indexSubcarrier\in\{\lowerFrequency,\mydots,\lowerFrequency+\numberOfShifts/\upsampleFactor-1\}$, where $\receivedSubcarrier[\indexSubcarrier]$ is the symbol on the $\indexSubcarrier$th subcarrier. After equalizing $\{\receivedSubcarrierCombined[\indexSubcarrier]\}$ with \ac{MMSE}-\ac{FDE}, by considering the scaled noise,   $\signalFactor$ can be calculated as $\signalFactor=\left(\upsampleFactor/\numberOfShifts\sum_{\indexSubcarrier=\lowerFrequency}^{\lowerFrequency+\numberOfShifts/\upsampleFactor-1} \frac{\FDSScoefnew[\indexSubcarrier]}{\FDSScoefnew[\indexSubcarrier]+1/\SNR}\right)^2$, where $\FDSScoefnew[\indexSubcarrier]=\sum_{\copyIndex=0}^{\upsampleFactor-1}|\fourierSeries[\indexSubcarrier+\copyIndex\numberOfShifts/\upsampleFactor]|^2$. Since $\FDSScoefnew[\indexSubcarrier]$ is a summation based on $\{\fourierSeries[\indexSubcarrier]\}$, the amplitude variation of $\FDSScoefnew[\indexSubcarrier]$ is less than the original case. Hence, the noise enhancement is substantially mitigated  at the expense of \ac{SE}  as shown in Section~\ref{sec:results}.


\def\numberOfUsers{N_{\rm user}}
\subsection{Compatibility with Other Features of Physical Layer}
Equation \eqref{eq:originalWaveform} provides insight into how existing techniques in \ac{OFDM}-based systems can be utilized for chirps.
\subsubsection{Multiple access} To support $\numberOfUsers$ users in the uplink within the same band without distorting the chirps, one simple approach is to up-sample the shaped sequence by a factor of $\numberOfUsers-1$ in the frequency domain. Since the up-sampling in the frequency domain causes repetition in the time  \cite{Frank_2008}, the transmitted signal  consists of $\numberOfUsers$ repetitions of \acp{CSC} within $\symbolDuration$. The zeroes due to the up-sampling in the frequency domain can be utilized by the other users, e.g., each user maps the up-sampled shaped sequence to a distinct set of subcarriers by shifting the sequence. If the channel is assumed to be flat (e.g., a strong line-of-sight path exists), the zeroes bins mentioned in Section~\ref{subsubsec:noiseEnc} can be also be utilized for multiple access.
\subsubsection{Multiple antennas} Equation \eqref{eq:originalWaveform} shows that if a \ac{MIMO} precoder is applied per subcarrier or a subcarrier group, it can distort the chirps. Hence, a \ac{MIMO} precoder can be applied based on the bandwidth of the chirp to avoid distortion. The diversity combining techniques, such as maximum-ratio combining, can also be trivially employed since the received signal is processed in the frequency domain.
\subsubsection{Channel estimation} Since the proposed framework utilizes \ac{CP}, it allows the receiver to estimate the channel in the frequency domain. In one approach, a single data symbol can be activated, e.g., $\dataSymbols[0]=1$ and $\dataSymbols[\indexSubcarrier\neq0]=0$. In this case, the elements of the shaped sequence can be utilized as pilots in the frequency domain. In another approach, the shaped sequence is upsampled and every other subcarrier is utilized for a pilot symbol as in a typical \ac{OFDM} transmission.
\subsubsection{DFRC} 
Achieving communications and radar functionality with the same waveform can address the under-utilized radar spectrum  \cite{Paul_2017}. The proposed framework enables that the correlation properties of the chirps can be exploited  while allowing them to carry information. For example, if only a portion of data symbols are activated with an index modulation \cite{basar_2013}, a large amount of information can be transmitted while utilizing the correlation properties of the chirps for range and velocity estimation.


\section{Analytical FDSS Filter Design}
In general, it is not trivial to obtain the \ac{FDSS} coefficients analytically for an arbitrary chirp.  In this section, we first discuss the chirps where the corresponding \ac{FDSS} coefficients can be expressed in closed-form. We then derive a theoretical framework for \ac{FDSS} coefficients to synthesize  chirp  with an arbitrary trajectory in time and frequency plane.

\subsection{FDSS Filters in Closed-Form for Special Chirps}

\subsubsection{Sinusoidal Chirps}
Let the instantaneous frequency of  $\basisFunction[{\amountOfShift[0]}]$ around the carrier frequency $\fcarrier$ be a sinusoidal function given by
$
\instantaneousFrequency[0]=\frac{\numberOfOccupiedSubcarriers}{2\symbolDuration}\cos\left({2\pi \frac{\timeVar}{\symbolDuration}}\right).
$
Therefore, $\angleSignal[0]=\frac{\numberOfOccupiedSubcarriers}{2}\sin\left({2\pi \frac{\timeVar}{\symbolDuration}}\right)$. It is well-known that $\constante^{\constantj\angleSignal[0]}$ can be decomposed as \cite{proakisfundamentals}
\begin{align}
\constante^{\constantj\frac{\numberOfOccupiedSubcarriers}{2}\sin\left({2\pi \frac{\timeVar}{\symbolDuration}}\right)}&=\sum_{\indexSubcarrier=-\infty}^{\infty}\besselFunctionFirstKind[\indexSubcarrier][\frac{\numberOfOccupiedSubcarriers}{2}]\constante^{\constantj2\pi \indexSubcarrier \frac{\timeVar}{\symbolDuration}}~,
\label{eq:bessel}
\end{align}
where $\besselFunctionFirstKind[\indexSubcarrier][\cdot]$ is the Bessel function of the first kind of order $\indexSubcarrier$. Hence, 
$\fourierSeries[\indexSubcarrier]$ is equal to $\besselFunctionFirstKind[\indexSubcarrier][\frac{\numberOfOccupiedSubcarriers}{2}]$ for sinusoidal chirps.

\subsubsection{Linear Chirps}
Assume that the instantaneous frequency of  $\basisFunction[{\amountOfShift[0]}]$ around the carrier frequency $\fcarrier$ changes form $-\frac{\numberOfOccupiedSubcarriers}{2\symbolDuration}$ Hz to $\frac{\numberOfOccupiedSubcarriers}{2\symbolDuration}$ Hz, i.e.,
$
\instantaneousFrequency[0]=\frac{\numberOfOccupiedSubcarriers}{2\symbolDuration}\left(\frac{2\timeVar}{\symbolDuration}-1\right),
$
which results in $\angleSignal[0]=\frac{\pi\numberOfOccupiedSubcarriers}{\symbolDuration}\left({ \frac{\timeVar^2}{\symbolDuration}-\timeVar}\right)$.
The shaping coefficients
$\fourierSeries[\indexSubcarrier]$ can  be obtained as 
$
\fourierSeries[\indexSubcarrier] = \sqrt{\frac{\pi}{\numberOfOccupiedSubcarriers}}\constante^{-\constantj\frac{(2\pi\indexSubcarrier)^2}{2\numberOfOccupiedSubcarriers}-\constantj\pi \indexSubcarrier} (\fresnelC[{\linearXone}] \texttt{+} \fresnelC[{\linearXtwo}]\texttt{+} \constantj\fresnelS[{\linearXone}] \texttt{+} \constantj\fresnelS[{\linearXtwo}])
$,
where $\fresnelC[{\cdot}]$ and  $\fresnelS[{\cdot}]$ are the Fresnel integrals with cosine and sine functions, respectively, and $\linearXone=(\numberOfOccupiedSubcarriers/2+2\pi\indexSubcarrier)/\sqrt{\pi\numberOfOccupiedSubcarriers }$ and $\linearXtwo=(\numberOfOccupiedSubcarriers/2-2\pi\indexSubcarrier)/\sqrt{\pi\numberOfOccupiedSubcarriers }$  \cite{cook1993radar}. 

\def\variableFunction{x}
\def\arbitrarFunction[#1]{f(#1)}
\def\fourierCoefa[#1]{a_{#1}}
\def\fourierCoefb[#1]{b_{#1}}
\def\indexFourierSum{n}
\def\eleSin[#1][#2]{\beta_{#1#2}}
\def\eleCos[#1][#2]{\alpha_{#1#2}}
\def\indexSequenceEle{m}
\def\kroneckerDelta[#1]{\delta\left(#1\right)}
\newcommand{\Conv}{%
	\mathop{\scalebox{1.5}{\raisebox{-0.2ex}{$\circledast$}}
	}
}
\subsection{FDSS Filters in Analytic-Form for Arbitrary Chirps}
Let $\arbitrarFunction[{\variableFunction}]$ be a periodic function with the period of $2\pi$, where $|\max{d\arbitrarFunction[\variableFunction]/d\timeVar}|=|\min{d\arbitrarFunction[\variableFunction]/d\timeVar}|=1$. For $\angleSignal[0]=\frac{\numberOfOccupiedSubcarriers}{2}\arbitrarFunction[{2\pi\frac{\timeVar}{\symbolDuration}}]$, $\instantaneousFrequency[0]$ changes between $-\numberOfOccupiedSubcarriers/2\symbolDuration$ and  $\numberOfOccupiedSubcarriers/2\symbolDuration$ Hz. By using the Fourier series of $\arbitrarFunction[{\variableFunction}]$, we can express $\constante^{\constantj\angleSignal[0]}$ as
\begin{align}
\constante^{\constantj\frac{\numberOfOccupiedSubcarriers}{2}\arbitrarFunction[{2\pi\frac{\timeVar}{\symbolDuration}}]}&=\constante^{\constantj\frac{\numberOfOccupiedSubcarriers}{2}\left(\frac{\fourierCoefa[0]}{2} + \sum_{\indexFourierSum=1}^{\infty}\fourierCoefa[\indexFourierSum]\cos(2\pi\frac{\indexFourierSum\timeVar}{\symbolDuration}) + \sum_{\indexFourierSum=1}^{\infty}\fourierCoefb[\indexFourierSum]\sin(2\pi\frac{\indexFourierSum\timeVar}{\symbolDuration})\right)}~.
\label{eq:arbitary}
\end{align}
Let $\eleCos[\indexFourierSum][\indexSubcarrier]$ and $\eleSin[\indexFourierSum][\indexSubcarrier]$  be the $\indexSubcarrier$th Fourier coefficients of   $\constante^{\constantj\frac{\numberOfOccupiedSubcarriers}{2}\fourierCoefb[\indexFourierSum]\cos(2\pi\frac{\indexFourierSum\timeVar}{\symbolDuration})}$ and $\constante^{\constantj\frac{\numberOfOccupiedSubcarriers}{2}\fourierCoefb[\indexFourierSum]\sin(2\pi\frac{\indexFourierSum\timeVar}{\symbolDuration})}$, respectively. By using \eqref{eq:bessel}, $\eleCos[\indexFourierSum][\indexSubcarrier]$ and $\eleSin[\indexFourierSum][\indexSubcarrier]$  can be obtained as
\begin{align}
\eleCos[\indexFourierSum][\indexSubcarrier] = \sum_{\indexSequenceEle=-\infty}^{\infty}\besselFunctionFirstKind[\indexSubcarrier][\frac{\fourierCoefa[\indexFourierSum]\numberOfOccupiedSubcarriers}{2}] \constantj^{\indexSubcarrier} \kroneckerDelta[\indexSequenceEle - \frac{\indexSubcarrier}{\indexFourierSum}]~,
\end{align}
and
\begin{align}
\eleSin[\indexFourierSum][\indexSubcarrier]= \sum_{\indexSequenceEle=-\infty}^{\infty}\besselFunctionFirstKind[\indexSubcarrier][\frac{\fourierCoefb[\indexFourierSum]\numberOfOccupiedSubcarriers}{2}]  \kroneckerDelta[\indexSequenceEle - \frac{\indexSubcarrier}{\indexFourierSum}]~,
\end{align}
respectively, where $\kroneckerDelta[\cdot]$ is Kronecker delta function. Since \eqref{eq:arbitary} is a multiplication of $\constante^{\constantj\frac{\numberOfOccupiedSubcarriers}{2}\fourierCoefb[\indexFourierSum]\cos(2\pi\frac{\indexFourierSum\timeVar}{\symbolDuration})}$ and $\constante^{\constantj\frac{\numberOfOccupiedSubcarriers}{2}\fourierCoefb[\indexFourierSum]\sin(2\pi\frac{\indexFourierSum\timeVar}{\symbolDuration})}$ for $\indexFourierSum=1,2,\mydots,\infty$, the  Fourier series of $\constante^{\constantj\frac{\numberOfOccupiedSubcarriers}{2}\arbitrarFunction[{\frac{\timeVar}{\symbolDuration}}]}$ can be calculated via the convolution theorem as
\begin{align}
\fourierSeries[\indexSubcarrier] =& \constante^{\constantj\frac{\numberOfOccupiedSubcarriers}{2}\frac{\fourierCoefa[0]}{2}} (\eleSin[\indexFourierSum=1,][\indexSubcarrier] \circledast \eleCos[\indexFourierSum=1,][\indexSubcarrier])\circledast(\eleSin[\indexFourierSum=2,][\indexSubcarrier] \circledast \eleCos[\indexFourierSum=2,][\indexSubcarrier])\circledast\mydots \nonumber \\=& \constante^{\constantj\frac{\numberOfOccupiedSubcarriers}{2}\frac{\fourierCoefa[0]}{2}}\Conv_{\indexFourierSum=1}^{\infty} (\eleSin[\indexFourierSum][\indexSubcarrier] \circledast \eleCos[\indexFourierSum][\indexSubcarrier])~,
\label{eq:manyconv}
\end{align}
Note that \eqref{eq:manyconv} requires  the  calculation of  infinitely many convolutions of infinite-length sequences. Although this is not tractable in general, for practical chirps, it can be  accurately calculated since 1) Fourier series quickly converge (i.e., limited $\indexFourierSum$) for  chirps used in practice, e.g., triangular chirps, 2) Due to the properties of  the Bessel function of the first kind, the limits of $\eleCos[\indexFourierSum][\indexSubcarrier]$ and $\eleSin[\indexFourierSum][\indexSubcarrier]$ are zeros as $|\indexSubcarrier|$ approaches to infinity, and 3) \ac{FDSS} filter design is often an offline procedure.

\subsubsection{Triangular Chirp}
Triangular chirp allows a radar system to estimate not only the range but also the target's velocity \cite{Saponara_2019}. The corresponding \ac{FDSS} filter for triangular chirp can be obtained   by using \eqref{eq:manyconv}, analytically. Assume that the down-chirp first is transmitted in the first half of the symbol with a following up-chirp. The corresponding $\arbitrarFunction[{\variableFunction}]$ within one period can be expressed as
\begin{align}
\arbitrarFunction[{\variableFunction}]= 
\begin{cases}
\frac{1}{\pi}\variableFunction^2+\variableFunction, 		& \text{if}~-\pi\le\variableFunction<0\\
-\frac{1}{\pi}\variableFunction^2+\variableFunction,     	& \text{if}~0\le\variableFunction<\pi
\end{cases}~,
\end{align}
The Fourier series coefficients of $\arbitrarFunction[{\variableFunction}]$ can then be obtained as $\fourierCoefa[\indexFourierSum]=0$ and
\begin{align}
\fourierCoefb[\indexFourierSum]= \frac{4-2\pi\indexFourierSum\sin(\pi\indexFourierSum)-4\cos(\pi\indexFourierSum)}{\pi^2\indexFourierSum}~.
\label{eq:bn}
\end{align}
If the up-chirp is transmitted on the first half with a following down-chirp, $\fourierCoefa[\indexFourierSum]=0$ and the corresponding $\fourierCoefb[\indexFourierSum]$ is equal to the negative of \eqref{eq:bn}.
Since  $\fourierCoefa[\indexFourierSum]=0$ for both cases, the  \ac{FDSS}  coefficients can be obtained analytically as $\fourierSeries[\indexSubcarrier] =  \Conv_{\indexFourierSum=1}^{\infty} \eleSin[\indexFourierSum][\indexSubcarrier]$.


\section{Numerical Results}\label{sec:results}
\begin{figure}[t]
	\centering
	\subfloat[FDSS: None (plain \ac{DFT-s-OFDM}).]{\includegraphics[width =1.75in]{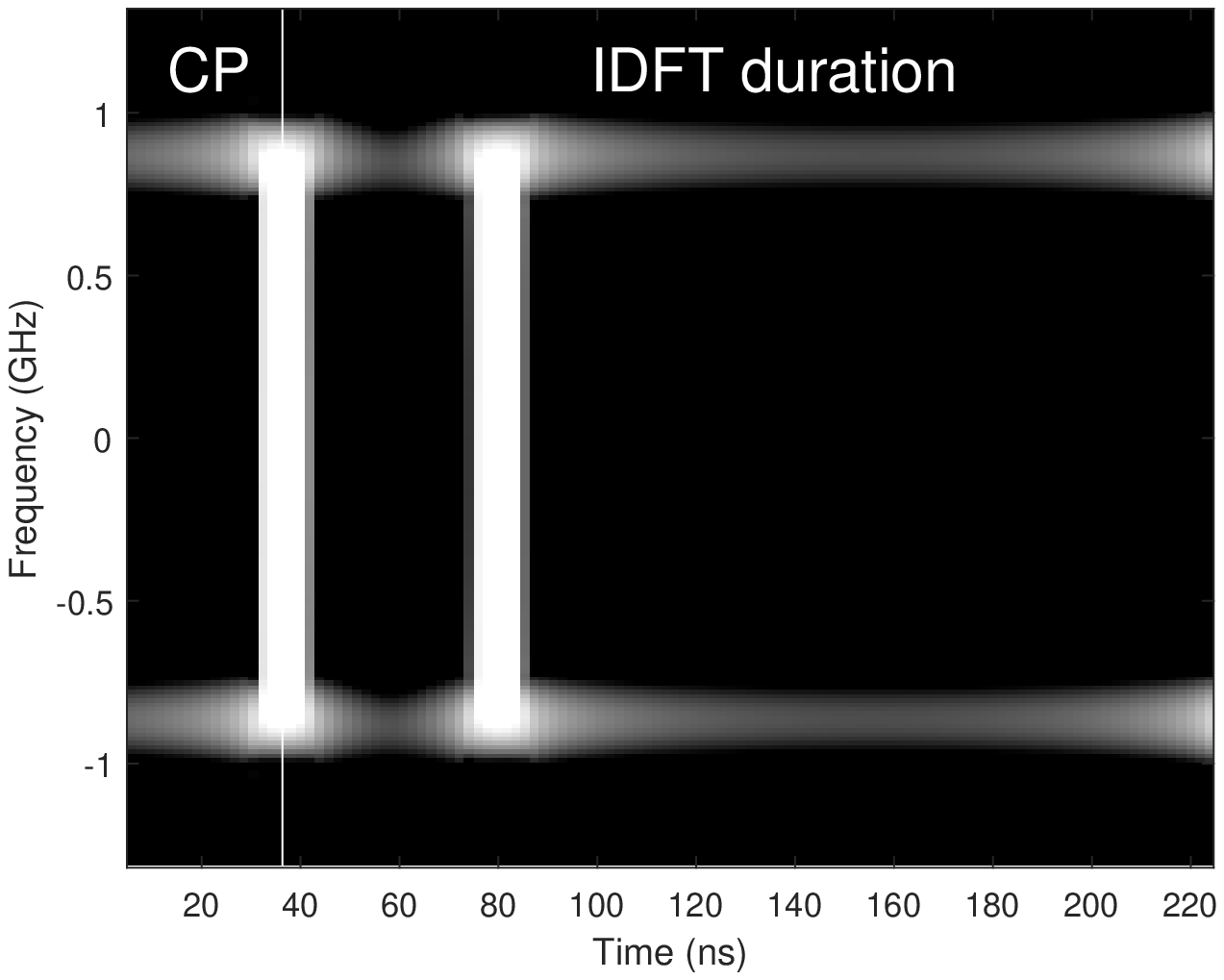}\label{subfig:plain}}	
	\subfloat[FDSS for linear chirps.]{\includegraphics[width =1.75in]{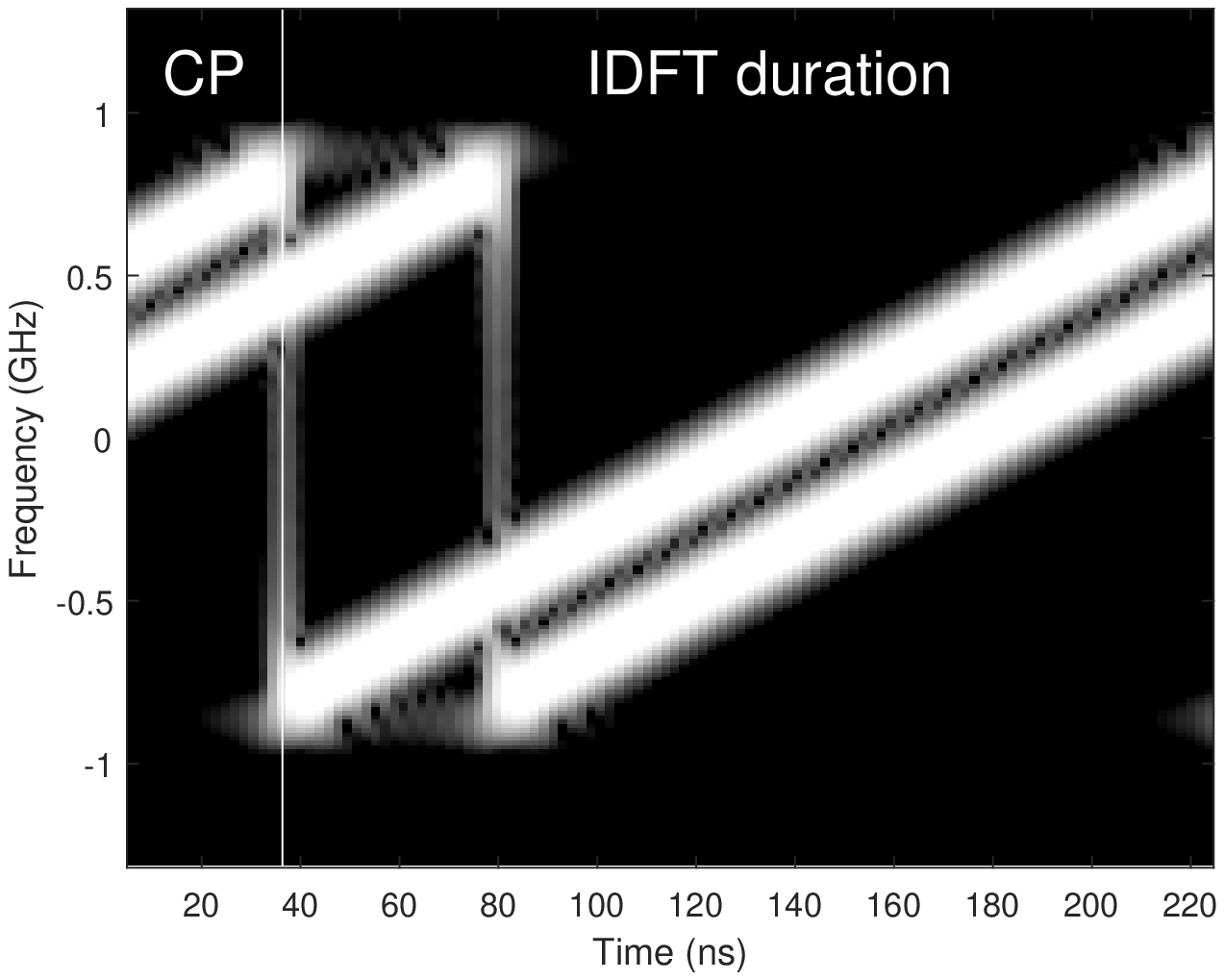}\label{subfig:linear}}\\
	\subfloat[FDSS for sinusoidal chirps.]{\includegraphics[width =1.75in]{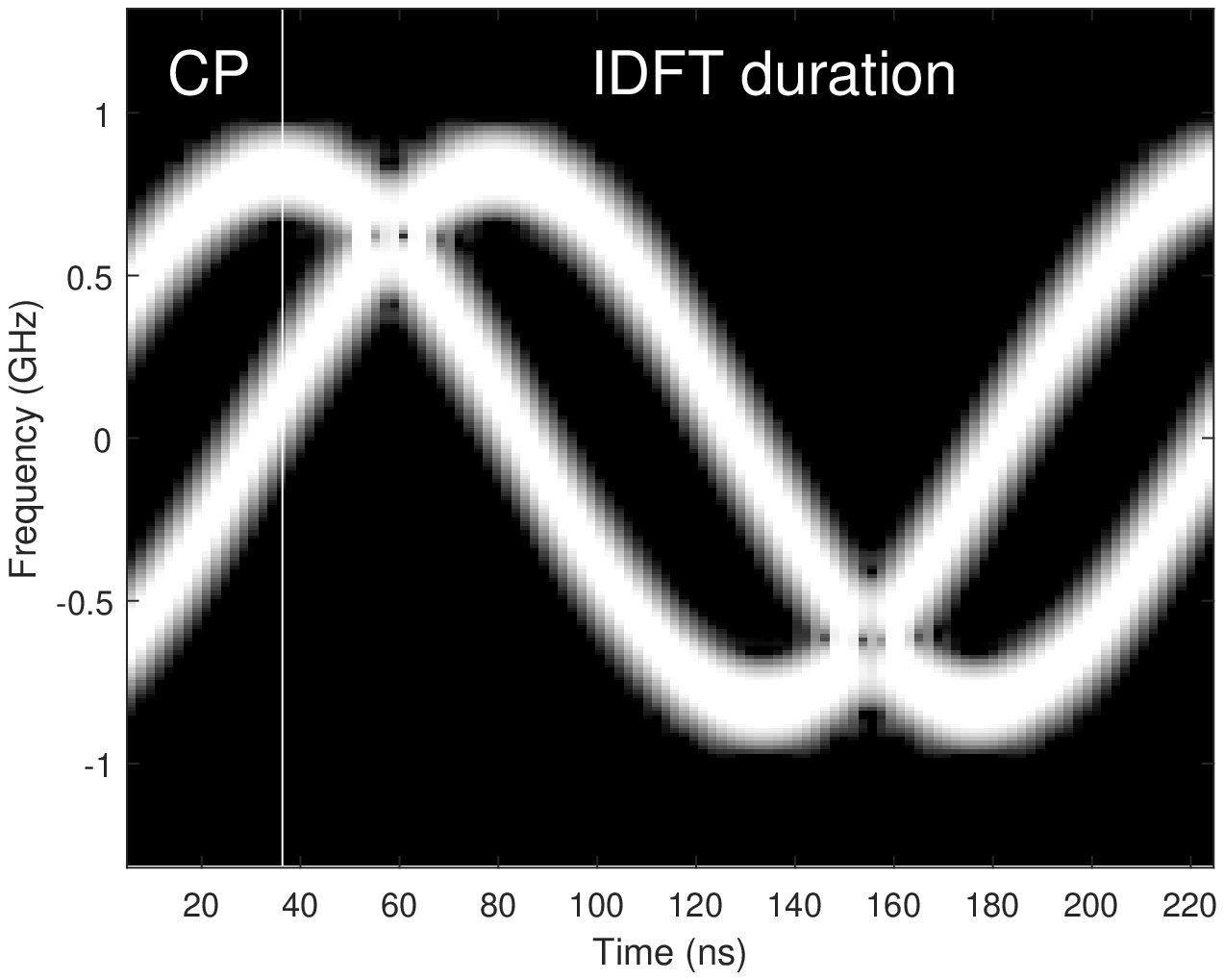}\label{subfig:sin}}
	\subfloat[FDSS for triangular chirps.]{\includegraphics[width =1.75in]{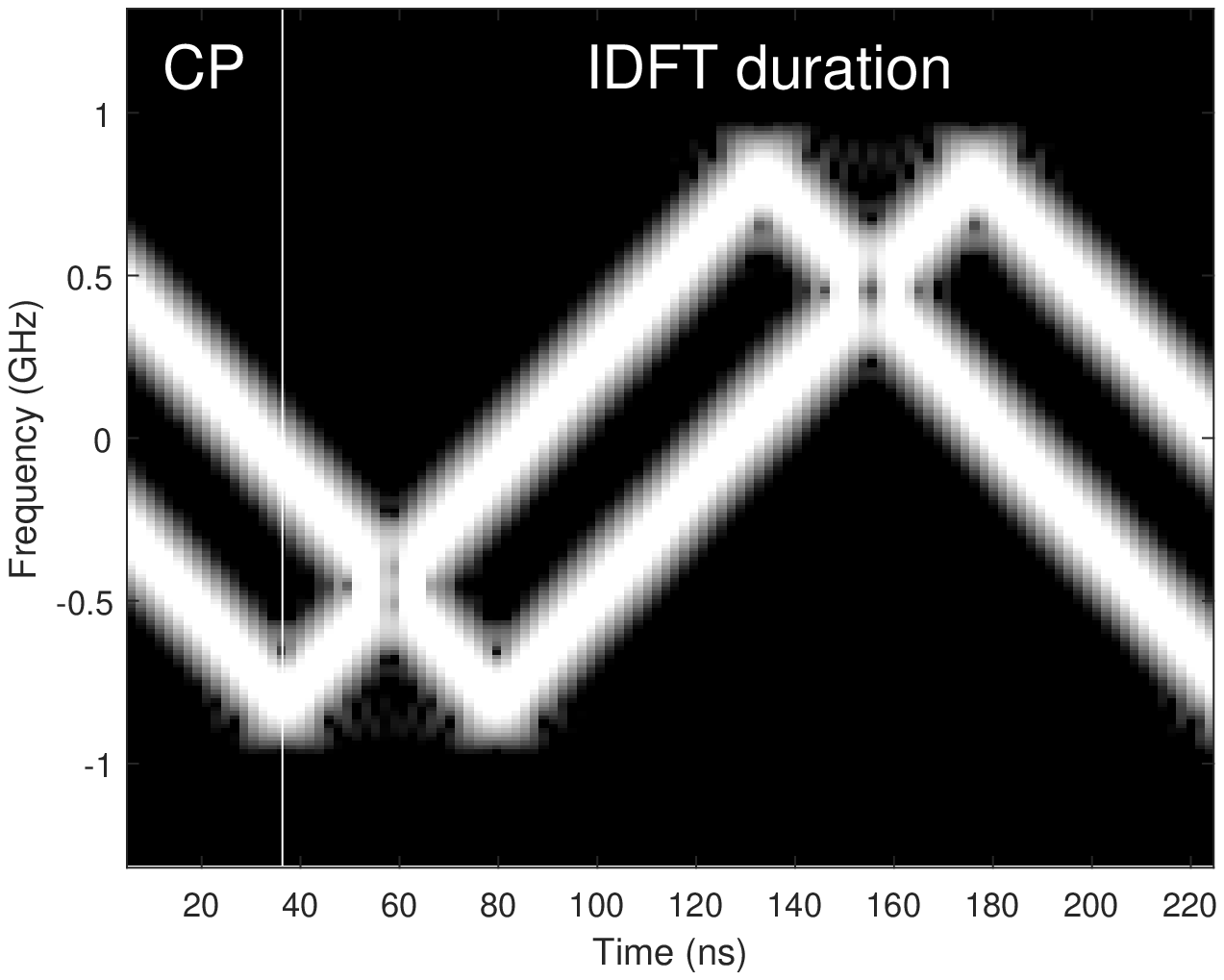}	\label{subfig:tri}}	
	\caption{\ac{DFT-s-OFDM} with various FDSSs can generate arbitrary chirps ($\dataSymbols[0]=\dataSymbols[75]=1$ and the rest of the symbols are set to zeros).}
	\label{fig:chirps}
	\vspace{-2mm}
\end{figure}
For simulations, we consider IEEE 802.11ay \ac{OFDM} PHY, where the symbol duration and the \ac{CP} duration are $\symbolDuration = 193.4$~ns and $\CPDuration=36.3$~ns, respectively.  We assume that the transmitter uses $\numberOfShifts=336$ subcarriers.  For chirps, we choose $\numberOfOccupiedSubcarriers=318$ to not distort chirps due to the truncation. We generate the data symbols based on \ac{QPSK}  and use $\numberOfShifts=336$ \acp{CSC}  unless otherwise stated. We consider a multipath channel where the power delay profile is $\{0, -10, -20\}$ dB. While the first path follows Rician distribution with a $K$-factor of 10, the other two paths are based on Rayleigh distributions. For the channel coding, IEEE 802.11ay \ac{LDPC} code  with the rate of $1/2$ is employed, where the codeword length is $672$. We consider linear, sinusoidal, and triangular \acp{CSC}. As a reference, we also provide the results for plain \ac{DFT-s-OFDM} (i.e., no \ac{FDSS}). For triangular chirp, we obtain the  \ac{FDSS} filter based on \eqref{eq:manyconv}. We assume that \ac{FDSS} filter is known at the receiver.

In \figurename~\ref{fig:chirps}, we demonstrate that a \ac{DFT-s-OFDM} transmitter is capable of synthesizing \acp{CSC} with arbitrary trajectories in time and frequency. For this analysis, we assume $\dataSymbols[0] = \dataSymbols[75]=1$ and the rest is set to 0. Since a plain \ac{DFT-s-OFDM} signal  is a form of single carrier waveform, the symbols $\dataSymbols[0] $ and $ \dataSymbols[75]$  appear as two pulses in time as in \figurename~\ref{fig:chirps}\subref{subfig:plain}. In contrast, the same symbols result in two linear, sinusoidal, and triangular \acp{CSC} transmitted in simultaneously as in \figurename~\ref{fig:chirps}\subref{subfig:linear}-\ref{fig:chirps}\subref{subfig:tri}, respectively. We  observe that there is a sudden frequency change for linear chirps. This is due to the fact that sinusoidal chirp is a  continuous periodic function while a linear chirp is a {\em discontinuous} periodic function in \eqref{eq:basisDecompose}. This abrupt change in the frequency can also be observed as a distortion that affects the constant-envelope nature of the linear chirp in the time. Theoretically, such distortion is inevitable for $\{\fourierSeries[\indexSubcarrier]\}$ with finite support. However, it can be mitigated by decreasing the maximum frequency deviation or allowing a tolerable amount of leakage beyond the channel bandwidth by choosing $\lowerFrequency<\floor{\numberOfShifts/2}-\numberOfShifts+1$ and $\upperFrequency>\floor{\numberOfShifts/2}$ in \eqref{eq:chirpWave}. The abrupt instantaneous frequency changes are avoided for the sinusoidal and triangular chirps. \figurename~\ref{fig:chirps}\subref{subfig:tri} also demonstrates that the triangular chirps can be  synthesized accurately with \eqref{eq:manyconv}.

\begin{figure}[t]
	\centering
	{\includegraphics[width =3.3in]{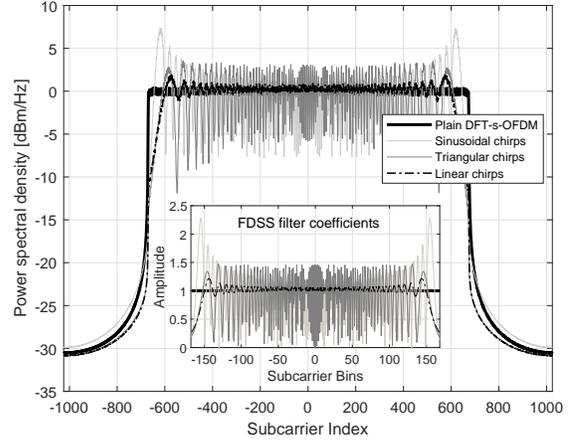}
	}
	\vspace{-2mm}
	\caption{Power spectral density and FDSS filter coefficients.}
	\vspace{-3mm}
	\label{fig:spectrum}
\end{figure}
In  \figurename~\ref{fig:spectrum}, the power spectral density and the \ac{FDSS}  coefficients of the aforementioned signals are compared. The main lobe of the spectrum is not flat for chirps. Particularly, a majority of the symbol energy is carried over the edge subcarriers for sinusoidal chirps. This is due to the fact that \ac{FDSS} for chirps does not distribute symbol energy to the subcarriers evenly. While the amplitude variations for linear chirps are relatively mild, they can be large for sinusoidal and triangular chirps. As discussed in Section~\ref{subsec:receiver}, large ripples degrade the \ac{BER} performance due to the noise enhancement with \ac{MMSE}-\ac{FDE}. Particularly, the magnitude of the shaping coefficients can be very small for sinusoidal and triangular chirps (see the fluctuations in  \figurename~\ref{fig:spectrum}). Therefore, the subcarriers with small magnitude shaping coefficients are more prone to noise. 

\begin{figure}[t]
	\centering
	{\includegraphics[width =3.3in]{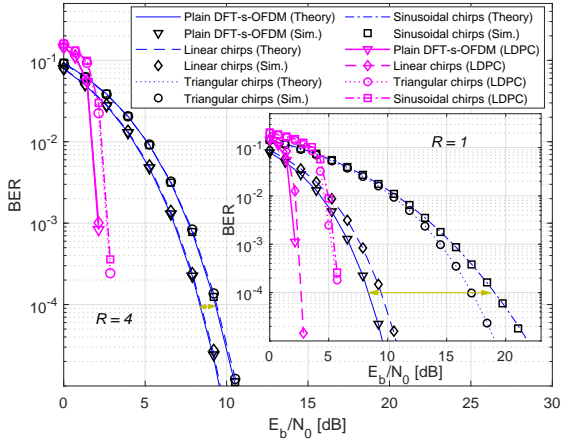}
	} 
	\vspace{-2mm}
	\caption{Coded and uncoded BER performance in AWGN for $\upsampleFactor=\{1,4\}$.}
	\vspace{-3mm}
	\label{fig:BERawgn}
\end{figure}
\begin{figure}[t]
	\centering
	{\includegraphics[width =3.3in]{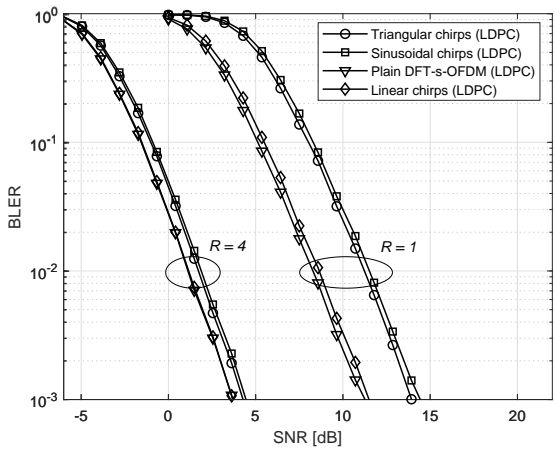}
	}
	\vspace{-2mm}
	\caption{BLER performance in fading channel  for $\upsampleFactor=\{1,4\}$.}
	\vspace{-3mm}
	\label{fig:BLERfading}
\end{figure}
In \figurename~\ref{fig:BERawgn} and \figurename~\ref{fig:BLERfading},  \ac{BER} and \ac{BLER} curves are provided. For both $\upsampleFactor=\{1,4\}$, the simulation results match well with the theoretical \ac{BER} curves based on  \eqref{eq:SNRpost}. For $\upsampleFactor=1$ and  uncoded bits, a large degradation  occurs for the sinusoidal and triangular chirps in  \ac{AWGN} while it is approximately 1 dB for linear chirps, as compared to plain \ac{DFT-s-OFDM}. The main reason for this degradation is non-unimodular \ac{FDSS}  coefficients. When \ac{LDPC} is introduced, for both \ac{AWGN} and fading channels, the degradation is approximately 0.5 dB for linear chirps while it is around 3~dB  for the sinusoidal and triangular chirps. 
For $\upsampleFactor=4$, the error rate substantially reduces for chirps as the receiver combines the signal in the frequency as discussed in Section~\ref{subsubsec:noiseEnc} and mitigate the noise enhancement. While the performance difference between \ac{DFT-s-OFDM} and linear chirps is negligible in this scenario, the  gap reduces to 0.8 dB for sinusoidal and triangular chirps.

%


\section{Concluding Remarks}\label{sec:conclusion}
In this study, we show that modulated \acp{CSC} can be generated with \ac{DFT-s-OFDM} via a well-designed \ac{FDSS} filter. For obtaining \ac{FDSS} filter analytically for an arbitrary chirp, we also develop a theoretical framework based on the Bessel function of the first kind and the Fourier series of the trajectory in time and frequency. We  show that a typical \ac{DFT-s-OFDM} receiver with a single-tap \ac{FDE}-\ac{MMSE} can decode the modulated chirps. We theoretically quantify the impact of the magnitude variations in the shaping coefficients on \ac{SNR} after the equalization and discuss how to mitigate noise enhancement through repetitions in the frequency. 
As the corresponding \ac{FDSS} for linear chirps has fewer magnitude variations as compared to the sinusoidal and triangular chirps, linear \acp{CSC} performs similar to the plain \ac{DFT-s-OFDM}.


The main benefit of the proposed approach is that it provides insight into how chirps can be synthesized and used for communications without introducing major modifications to the physical layer of today's wireless standards. Since the techniques developed for \ac{OFDM} can  be utilized through the introduced framework, it paves the wave for developing new methods for \ac{DFRC}  and \ac{IoT} applications.

\bibliographystyle{IEEEtran}
\bibliography{bandlimitedChirps}

\end{document}